\documentstyle[aps,twocolumn]{revtex}
\input psfig
\begin{document}
\draft
\title{Composite fermions in the Fractional Quantum Hall Effect:
Transport at finite wavevector.}
\author{A.D. Mirlin$\sp{1,2}$  
 and  P. W\"{o}lfle$\sp{1}$}
\address{
$\sp{1}$ Institut f\"{u}r Theorie der Kondensierten Materie,
  Universit\"{a}t Karlsruhe, 76128 Karlsruhe, Germany} 
\address{
$^2$ Petersburg Nuclear Physics Institute, 188350 Gatchina, St.Petersburg, 
Russia.}
\date{\today}
\maketitle
\narrowtext
\tighten
\begin{abstract}
We consider the conductivity tensor for composite fermions in a close to
half-filled Landau band in the temperature regime where the scattering
off the potential and the trapped gauge field of random impurities
dominates. The Boltzmann equation approach is employed to calculate
the quasiclassical transport properties at finite effective magnetic
field, wavevector and frequency. We present an exact solution of the
kinetic equation for all parameter regimes. Our results allow a
consistent description of recently observed surface acoustic wave
resonances and other findings.

\end{abstract}

The properties of two-dimensional (2D) electron systems in high
magnetic fields appear to be well described by the model of composite
fermions (CF's) \cite{jain,hlr}. It postulates the existence of
fermionic quasiparticles consisting of electrons to which an even
number of flux quanta has been attached. The corresponding mapping can
be expressed by introduction of a Chern-Simons (CS) gauge field
\cite{hlr,lofra}. The average magnetic field $B_{\mbox{eff}}$ experienced 
by the CF's is the difference of the external magnetic field $B$ and
the mean field of the flux tubes, 
$B_{\mbox{fl}}=\tilde{\phi}\phi_0n$, where
$\phi_0=hc/e$ is the flux quantum, $\tilde{\phi}$ is the (even) number
of flux quanta per electron and $n$ is the electron density. The CF
model appears to capture the effect of the Coulomb interaction in
maintaining optimum electron separation particularly near even
denominator fillings of the Landau level, $\nu=1/(2q)$, when choosing
$\tilde{\phi}=2q$ ($q=1,2,\ldots$). Then the effective magnetic field
$B_{\mbox{eff}}$ is largely reduced near and exactly zero at these
filling factors. In this regime the CF model implies the existence of
a Fermi surface with wave number $k_F=\sqrt{4\pi n}$ in a weak
magnetic field \cite{hlr}. At low temperatures the CF's are subject to
elastic impurity scattering off the random potential created by a
remote layer of donors and, more importantly, off the frozen random
gauge field configuration induced by the impurities. In addition, the
CF's interact via the scalar (Coulomb) potential and the CS
gauge field. 

The existence of new particles with a definite effective mass moving
under the influence of an effective magnetic field $B_{\mbox{eff}}$
has been demonstrated in a number of experiments. For example, the
interpretation of magnetooscillations in the longitudinal resistivity
in terms of the Shubnikov--de Haas (SdH) effect for CF's has been very
successful \cite{cf,aamw}. The observation of resonance effects for
the motion of charge carriers in the field of surface acoustic waves
(SAW) is another example \cite{willett}. However, a closer
examination suggested \cite{willett2} 
that  the experimental observation of these resonances
is inconsistent with the conventional interpretation in the range of
wave frequencies used. A quantity of central interest for the
theoretical description of this phenomenon is  the conductivity tensor
at finite wave vector $\sigma_{\mu\nu}(\omega,\bbox{q})$, the calculation
of which we will reexamine in the present paper. The
momentum-dependent conductivity $\sigma_{\mu\nu}(\omega,\bbox{q})$,
which determines the gauge field polarization operator,
is also of  importance for the damping of SdH oscillations
by static and dynamic fluctuations of the gauge field \cite{aamw,lee}.

In this letter we present a quasiclassical calculation of
$\sigma_{\mu\nu}(\omega,\bbox{q})$ in a magnetic field, taking into
account the impurity scattering of fermions
 off a static random magnetic field and random potential. Surprisingly,
despite the general character and the importance of the problem, we
did not find a correct calculation of
$\sigma_{\mu\nu}(\omega,\bbox{q})$ in the literature even for the usual
case of a short-range correlated random potential, except in the
collisionless limit \cite{simon}. The standard reference \cite{cohen}
appears to have missed the fact that, since at finite wave vector
$\bbox{q}$ the current and density fluctuations are coupled,
the relaxation time approximation in its simplest form cannot be
used because it violates the particle number conservation.
 In particular, the results of Ref.\cite{cohen} (used later
for interpretation of the SAW experiments in
Refs.\cite{hlr,willett,willett2}) violate the exact property
$\lim_{\omega\to 0}q_\mu\sigma_{\mu\nu}(\omega,\bbox{q})=0$, which is
a consequence of the current conservation. The particle number
conservation has to be taken into account properly in modelling the
collision integral of the Boltzmann equation, as shown below. Another
new aspect of the transport problem we are considering is the extreme
angular dependence of the scattering off a random magnetic field,
which is so strongly peaked in the forward direction, that the
relaxation time approximation is not adequate.

In quasiclassical transport theory the components of the electrical
current $j_\mu$ may be expressed in terms of the distribution function
$f(\bbox{k};\bbox{r},t)$ of fermions of momentum $\bbox{k}$ at
position $\bbox{r}$ and time $t$ as
$j_\mu(\bbox{r},t)=e\int(dk)(k_\mu/m^*)f(\bbox{k};\bbox{r},t)$,
where $e$ and $m^*$ are the charge and the effective mass of the
carriers. We are interested in the linear response
$f_1(\bbox{k};\bbox{r},t)$ to an applied electric field
$\bbox{E}(\bbox{r},t)=\bbox{E}\exp(-i\omega t+i\bbox{qr})$ in the
presence of a weak  magnetic field $\bbox{B_0}$ perpendicular to the
plane, such that $\omega_c\tau_s\ll 1$ (here $\omega_c=eB_0/m^*c$ is
cyclotron frequency and $\tau_s^{-1}$ is the width of the Landau
levels). The function $f_1$ is obtained from the kinetic equation
$$
i(\omega-\bbox{vq})f_1
+{e\over c}(\bbox{v}\times\bbox{B_0})\cdot\bbox{\nabla_k}f_1 
+e\bbox{E}\bbox{\nabla_k}f_0=C\{f_1\},
$$
where $f_0$ is the equilibrium distribution function,
$\bbox{v}=\bbox{k}/m^*$ (we assume isotropic Fermi surface) and
$C\{f_1\}$ is the collision integral. The external force
$\propto\bbox{Ev}(\partial f/\partial\epsilon)$ acts only on states at
the Fermi circle ($\epsilon=\epsilon_F$), and therefore $f_1$ is only a
function of the angle $\phi$ of $\bbox{k}$ with the $\hat{x}$ axis,
say. In terms of the angular momentum eigenfunctions
$\chi_m(\phi)=e^{im\phi}$, the collision integral is given as
\begin{equation}
C\{f_1\}=\sum_{m=-\infty}^{\infty}\lambda_m
\int_0^{2\pi}{d\phi'\over 2\pi}\exp\{im(\phi-\phi')\}f_1(\phi').
\label{e2}
\end{equation}
Particle number conservation requires $\lambda_0=0$, also
$\lambda_m=\lambda_{-m}=\lambda_m^*$, since $C$ is a hermitean and
non-negative operator. The usual relaxation time approximation used in
Ref.\cite{cohen}, $C\simeq f_1/\tau$, is equivalent to setting
$\lambda_m=1/\tau$ for all $m$, obviously violating the requirement
$\lambda_0=0$. This can be corrected by approximating 
\begin{equation}
C\simeq (1/\tau)[f_1-\int f_1 d\phi/2\pi],
\label{e2a}
\end{equation}
implying $\lambda_0=0$ and
$\lambda_m=1/\tau$ for all $|m|\ge 1$ (in the following called model
I). Model I describes isotropic impurity scattering and
is applicable for short-range correlated random potential.
For scattering off a random magnetic field, when the transition rate
$W(\phi-\phi')$ for scattering of a particle from angle $\phi'$ to
angle $\phi$ on the Fermi circle is given (in Born approximation)
by \cite{amw}
$W(\phi-\phi')=\tau^{-1}\cot^2[(\phi-\phi')/2]$, model I is not
adequate. In this case, $W$ is seen to diverge in the forward
scattering limit $\phi=\phi'$. The eigenvalues of the collision
operator
\begin{equation}
C\{f_1\}=\int(d\phi'/2\pi)W(\phi-\phi')[f_1(\phi)-f_1(\phi')]
\label{e3}
\end{equation}
are then given for $m\ne 0$ 
by $\lambda_m=\tau^{-1}(2|m|-1)$, i.e. they increase
linearly with $|m|$ (to be called model II). 
The larger $|m|$, the deeper the forward
scattering divergence is probed.  Finally, if both random potential
and random magnetic field  scattering mechanisms are present,
we have $\lambda_m=\tau^{-1}(1-2p+2p|m|)$, with $0\le p\le 1$,
where $p=0$ and $p=1$ correspond to pure models I and II respectively.
In the sequel, we solve the Boltzmann equation exactly for arbitrary
values of the frequency $\omega$, momentum $\bbox{q}$ and magnetic
field $\bbox{B_0}$, and for arbitrary value of $p$.

{\it Solution of the kinetic equation.} We define a dimensionless
distribution function $g(\phi)$ by
$f_1(\bbox{k};\bbox{q},\omega)=(-\partial
f_0/\partial\epsilon)elEg(\phi)$, where $l=v\tau$ is the transport
mean free path. The function $g(\phi)$ obeys the integro-differential
equation
\begin{equation}
i\tau(\omega-vq\cos\phi)g-\omega_c\tau{\partial
g\over\partial\phi}-\cos(\phi_E-\phi)=\tau C\{g\},
\label{e4}
\end{equation}
where $\tan\phi_E=E_y/E_x$ and we have chosen the direction of the
momentum $\bbox{q}\parallel\hat{x}$. 
Expanding $g$ in terms of eigenfunctions
$\chi_m(\phi)$, $g(\phi)=\sum_{m=-\infty}^\infty g_m\chi_m(\phi)$, yields
the recursion relations
\begin{equation}
a_mg_m-b(g_{m+1}+g_{m-1})=S_m
\label{e5}
\end{equation}
with $a_m=-i\omega\tau+im\omega_c\tau+\lambda_m\tau$, $b=-ivq\tau/2$,
and $S_m=(1/2)\exp(-im\phi_E)\delta_{|m|,1}$. 
We note that the ``angular momentum'' components $g_m$ only mix
through the term proportional to $q$. 
For positive $m\ge 2$,
the system (\ref{e5}) can be solved to yield
\begin{equation}
g_2=R_+ g_1\ ;\qquad R_+=(b/a_2)P_3/P_2\ ,
\label{e6}
\end{equation}
where $P_k$ are power series in $b^2$ (thus, in $q^2$):
\begin{equation}
P_k=\sum_{n=0}^\infty(1/n!)(-b^2/\alpha_+)^n/
\prod_{l=k}^{n+k-1}a_l
\label{e7}
\end{equation}
Here $\alpha_+=\lim_{m\to\infty}(a_m/m)$, and we assume $a_m$ to
increase at most linearly with $m$ as $m\to\infty$; the condition
which is fulfilled for the models considered. For negative $m$ we find
analogously $g_{-2}=R_-g_{-1}$, where $R_-(\omega_c)=R_+(-\omega_c)$
and $\alpha_-(\omega_c)=\alpha_+(-\omega_c)$. In terms of the
components $g_1$, $g_{-1}$ of the distribution function the components
of the conductivity tensor are given by
$\sigma_{x\mu}=\sigma_0(g_1+g_{-1})$ and 
$\sigma_{y\mu}=i\sigma_0(g_1-g_{-1})$, where $\sigma_0=e^2n\tau/m^*$
and the index $\mu$ corresponds to the direction of the electric field
determined by the angle $\phi_E$. Using eq.(\ref{e6}) and its
counterpart for negative $m$, we solve the system of the remaining three
equations from (\ref{e5}) (with $m=0,\pm 1$) and find
\begin{eqnarray}
\sigma_{xx}&=&\sigma_0i\omega\tau(\gamma_++\gamma_-)/D\ ;\nonumber\\
\sigma_{xy}&=&\sigma_0\omega\tau(\gamma_--\gamma_+)/D=-\sigma_{yx}\ ;
\label{e8} \\
\sigma_{yy}&=&\sigma_0 [4b^2+i\omega\tau(\gamma_++\gamma_-)]/D\ , \nonumber
\end{eqnarray}
where $D/2=b^2(\gamma_++\gamma_-)+i\omega\tau\gamma_+\gamma_-$ and
$\gamma_{\pm}=1-i\omega\tau\pm i\omega_c\tau-bR_{\pm}$. For the chosen
direction of $\bbox{q}\parallel\hat{x}$ we find as expected
$\sigma_{x\mu}(q,\omega)\propto\omega$ for $\omega\to 0$, $q\ne 0$. The
exact solution (\ref{e8})
of the Boltzmann equation for any impurity collision integral 
can be easily evaluated for any finite value of $q$, since the power
series representation of the functions $P_k$, eq.(\ref{e7}), 
is rapidly converging. For the special cases of both models I and II
the eigenvalues of the collision integral $\lambda_m$ and consequently
the coefficients $a_m$ are linear function of $m$, so that the power 
series representation of $P_k$ is reduced to that of the Bessel function
$J_\nu(z)$, with complex valued index $\nu$ and argument $z$:
\begin{equation}
R_\pm=J_{2+\beta_\pm}(Q_\pm)/J_{1+\beta_\pm}(Q_\pm);
\ \  Q_\pm=-{ivq\tau / \alpha_\pm}\ ,
\label{e9}
\end{equation}
with $\alpha_\pm=2p\pm i\omega_c\tau$,
$\beta_\pm=(1-2p-i\omega\tau)/\alpha_\pm$.

Substituting eq.(\ref{e9}) in (\ref{e8}) and using the recursion
relation for the Bessel functions,
$z[J_{\nu+1}(z)+J_{\nu-1}(z)]=2\nu J_\nu(z)$, 
we finally get the conductivity tensor  for
arbitrary values of the three dimensionless parameters $Q=qR_c\equiv
qv/\omega_c$, $S=\omega_c\tau$, and $T=\omega/\omega_c$:
\begin{eqnarray}
&&\sigma_{xx}={(i\sigma_0/ QS)
(J_{-}J_{1+}-J_+J_{1-}) \over J_{-}J_+ -
(Q/2T)(J_{-}J_{1+}-J_+J_{1-})}; \nonumber \\
&&\sigma_{yx}=-\sigma_{xy}={(\sigma_0/ QS)
(J_{-}J_{1+}+J_+J_{1-}) \over J_{-}J_+ -
(Q/2T)(J_{-}J_{1+}-J_+J_{1-})}; \nonumber \\
&&\sigma_{yy}=\sigma_{xx}+{(2i\sigma_0/ ST)
J_{1-}J_{1+} \over J_{-}J_+ -
(Q/2T)(J_{-}J_{1+}-J_+J_{1-})}; \nonumber\\
&& J_{\pm}=J_{\pm [T+(1-2p)i/S]/(1\pm 2pi/S)}(Q/(1\pm 2pi/S));\nonumber \\
&& J_{1\pm}=J_{1\pm [T+(1-2p)i/S]/(1\pm 2pi/S)}(Q/(1\pm 2pi/S))
\label{e10}
\end{eqnarray}
In the limiting case $p=0$ (model I)  the formulas (\ref{e10})
simplify, since the arguments of all the Bessel functions are equal to
$Q$: $J_{\pm}=J_{\pm (T+i/S)}(Q)$, $J_{1\pm}=J_{1\pm (T+i/S)}(Q)$.
In this case eqs.(\ref{e10}) can be rewritten in a slightly modified
form (similar to that obtained in Ref.\cite{simon} in the limit
$\tau=\infty$) by using the Bessel functions identity 
$$
J_-J_{1+}-J_+J_{1-}={2(T+i/S)\over Q}\left[J_-J_+-{\sin\pi(T+i/S)\over
\pi(T+i/S)}\right]  
$$

For the particular case of model I, the kinetic equation (\ref{e4}),
(\ref{e2a}) can be also solved in a different way. With the notation
$\int(d\phi/2\pi)g(\phi)=g_0$,  eq.(\ref{e4}) reduces to a first order
differential equation, which has the solution
\begin{eqnarray}
&& g(\phi)=\int_{-\infty}^\phi d\phi'\exp\{K(\phi,\phi')\}
{g_0-\cos(\phi_E-\phi')\over \omega_c\tau}\ ; \label{e50}\\
&& K(\phi,\phi')=-{\phi-\phi'\over\omega_c}(\tau^{-1}-i\omega)-
{ivq\over\omega_c}(\sin\phi-\sin\phi')
\label{e51}
\end{eqnarray}
Calculating now $\int d\phi g(\phi)$, we get a selfconsistency
equation for $g_0$. Substituting the found value of $g_0$ back into
eq.(\ref{e50}), and calculating the conductivity components, we find
\begin{eqnarray}
&&  \sigma_{xx}=-2i\sigma_0{\omega\tau\over(qv\tau)^2}
\left(1-i\omega\tau{A_{00}\over A_{00}-\omega_c\tau}\right)\ ;
\nonumber \\
&&  \sigma_{yx}=-\sigma_{xy}=-2\sigma_0{\omega\over vq}
{A_{s0}\over A_{00}-\omega_c\tau} \ ; \nonumber \\
&&\sigma_{yy}=-{2\sigma_0\over\omega_c\tau}\left(
{A_{0s}A_{s0}\over A_{00}-\omega_c\tau} - A_{ss} \right)\ ;
\label{e52} \\
&& A_{ij}={1\over 2\pi}\int_0^{2\pi}d\phi \psi_i(\phi)
\int_{-\infty}^\phi d\phi' \psi_j(\phi') \exp\{K(\phi,\phi')\}\ ; \nonumber\\
&& \psi_0(\phi)=1\ ;\ \ \psi_s(\phi)=\sin\phi
\nonumber
\end{eqnarray}
In this form the results for model I are more suitable
for consideration of the zero magnetic field limit (see below).

{\it Limiting cases.} Despite their apparent simplicity, the results
(\ref{e10}) possess a reach variety of regimes with various types of
behavior depending on the relations between $S$, $Q$, and $T$; 
we consider only some of them below.
In the limit $q=0$ we have $R_\pm=0$, so that the
general results (\ref{e8}) reduce to the usual Drude formulas. In the
collisionless limit $\tau=\infty$, formulas (\ref{e10}) 
 can be proven to reduce to the results of Simon and
Halperin \cite{simon}. 

In zero magnetic field, $\omega_c=0$, we have $\sigma_{xy}=0$ and 
\begin{eqnarray}
&&
\sigma_{xx}^I=2i\sigma_o{\rho\over\kappa^2}
{(1-i\rho)(\sqrt{1+\kappa^2/(1-i\rho)^2}-1)\over
1-(1-i\rho)\sqrt{1+\kappa^2/(1-i\rho)^2} } \nonumber \\
&& \sigma_{yy}^I=2\sigma_0(\sqrt{1+\kappa^2/(1-i\rho)^2}-1)(1-i\rho)/\kappa^2
 \nonumber \\
&&\sigma_{xx}^{II}={2\sigma_0\over\kappa}\left[{I_{(-i\rho-1)/2}(\kappa)
\over
I_{(1-i\rho)/2}(\kappa)} - {\kappa\over i\rho}\right]^{-1} 
 \nonumber \\
&& \sigma_{yy}^{II}={2\sigma_0\over\kappa}
{I_{(1-i\rho)/2}(\kappa)\over
I_{(-i\rho-1)/2}(\kappa)}\,
 \label{e10a}
\end{eqnarray}
where the superscripts $I$ and $II$
refer to the models I and II respectively,
$\kappa=qv\tau$, $\rho=\omega\tau$, and $I_\nu(z)$ is the
modified Bessel function \cite{note2}. 
 In particular, in the large wavevector
limit, $\kappa\gg 1,\rho$, we get for both the models
\begin{equation}
\sigma_{xx}\simeq -ie^2\nu\omega/q^2\ ;\qquad 
\sigma_{yy}\simeq e^2v\nu/q\ ,
\label{e11}
\end{equation}
where $\nu=m^*/2\pi$ is the density of states at Fermi surface. These
results are identical to those in the collisionless limit. We conclude
that although the eigenvalues $\lambda_m$ of the collision operator $C$
of model II grow with $m$ without bound, for $qv\gg
\tau^{-1},\omega_c$ the collision integral
$C$ is negligible in comparison with $vq$. 

Finally, let us consider the case of relatively strong magnetic fields,
 $S\equiv\omega_c\tau\gg 1$, corresponding to the region where
most of the resonance and magnetooscillation effects for the composite
fermions are observed \cite{note}. For these phenomena the behavior of
 conductivity at low frequencies $\omega\ll \omega_c,qv$
and finite wavevectors $qR_c\gtrsim 1$ is important.
In this regime, $\sigma_{xx}$ and $\sigma_{xy}$ are small (since
 they vanish at $\omega=0$), and $\sigma_{yy}$ plays the most
 important role. We find 
\begin{eqnarray}
&&\sigma_{yy}^{I}\simeq 2\sigma_0 J_1^2(Q)/(1-J_0^2(Q)-i\omega\tau)\ ;
\nonumber\\
&&\sigma_{yy}^{II}\simeq\left\{
\begin{array}{ll}
\displaystyle{
{ \sigma_0 J_1^2(Q)/Q^2 \over J_1^2(Q)+J_0^2(Q)-J_0(Q)J_1(Q)/Q} }\ ,&\ 
Q/S\ll 1 \\
2\sigma_0/QS\ ,&\  Q/S\gg 1
\end{array} \right.
\nonumber
\end{eqnarray}
Note that for the model II at a large wavevector, $Q\gg S$, the
zero magnetic field result, eq.(\ref{e11}), is restored, whereas it is
not the case for the model I.

{\it SAW resonances.}
As an important example of application of our results, we will now
reconsider the interpretation of experimental data on the 
SAW resonances \cite{willett,willett2}. As discussed in
Ref.\cite{willett2}, using the formulas from Ref.\cite{cohen}
for the CF conductivity does not allow to explain the observation of
resonances at the experimental frequency $\omega=2\pi\times 10.7\
\mbox{GHz}$ if the value of the effective mass extracted from the SdH
measurements, $m^*\simeq 0.8 m_e$, is assumed ($m_e$ being the free
electron mass). We have, however,
recalculated the SAW velocity shift \cite{willett,willett2},
\begin{eqnarray}
&& \Delta v/v = (\alpha^2/2) 
\mbox{Re}[1+i\sigma_{xx}^{(e)}/\sigma_m]^{-1};\ \ 
 \alpha^2/2 =3.2\cdot 10^{-4}; \nonumber\\
&&\sigma_{xx}^{(e)}\simeq \rho_{yy}/\rho_{xy}^2;
\ \ \rho_{xy}=2h/e^2; \label{e13}\\
&&\rho_{yy}=\sigma_{xx}/(\sigma_{xx}\sigma_{yy}+\sigma_{xy}^2) 
\nonumber
\end{eqnarray}
 making use of the correct formulas for
$\sigma_{\mu\nu}(\omega,q)$  derived above and setting  $m^*= 0.8 m_e$.
The constant $\sigma_m$ in eq.(\ref{e13})
and the transport time $\tau$ were used as parameters to
optimize the fit to the experimental data from Fig.1 of
Ref.\cite{willett2}. We get very good fits to the experimentally
observed resonances by using any of the models I or II, or a mixed
model with combined mechanism of scattering (Fig.1), 
at $\sigma_m=1\cdot 10^6\ \mbox{cm/s}\simeq 10^{-6}\Omega^{-1}$, which
is close to expected values of this parameter \cite{willett,willett2}. 
Note that the value of the transport time $\tau$ found from optimizing
the fit depends appreciably upon the model assumed, varying from
$\tau= 20\mbox{ ps}$ to $\tau= 120\mbox{ ps}$ in the limiting cases of
 the models I and II respectively.

\vspace{-0.6cm}
\begin{figure}
\centerline{\psfig{figure=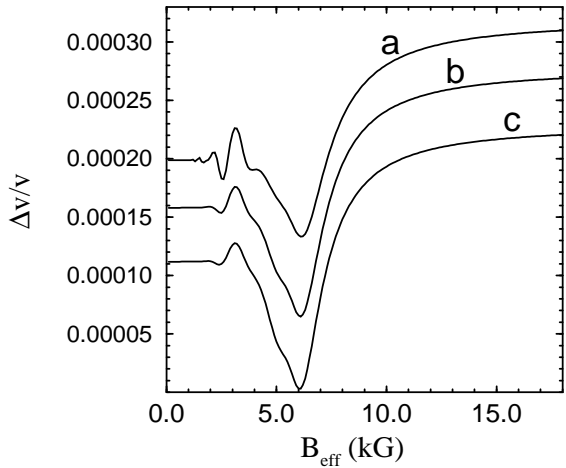,width=9cm}}
\vspace{-0.3cm}
\caption{Velocity shift of the  
surface acoustic waves, eq.(\protect\ref{e13}),  as a function of
the effective magnetic field experienced by composite fermions, with
to the following values of the parameters:
$\omega=2\pi\cdot 10.7\mbox{ GHz}$, 
$q=2\pi/(2.7\cdot 10^{-5}\mbox{ cm})$, 
$m^*=0.8m_e$, $n=1.6\cdot 10^{11}\mbox{ cm}^{-2}$,
and $\sigma_m=1\cdot 10^6\ \mbox{cm/s}$. 
The three curves shown correspond to:
a) model I ($p=0$), $\tau=20\mbox{ ps}$;
b) mixed model ($p=0.3$), $\tau=50\mbox{ ps}$;
c) model II ($p=1$), $\tau=120\mbox{ ps}$.
For clarity, the curves a) and b) are off-set by 0.0001 and 0.00005
respectively.
The principal and secondary resonances (minima) are seen at
$B_{\mbox{eff}}\simeq 6\mbox{ kG}$ and $B_{\mbox{eff}}\simeq 2.5\mbox{
kG}$, as observed in the experiment \protect\cite{willett2}. } 
\label{fig1}
\end{figure}

\vspace{-0.2cm}
In conclusion, we have presented a solution of a very general
problem: calculation of the semiclassical conductivity tensor of
fermions in a random potential and/or random magnetic field in 2D 
at finite values of frequency $\omega$ and momentum $q$ and in uniform
magnetic field $B_0$. We have pointed out that explicit account of the
particle number conservation in modelling  the collision integral is
crucially important at finite $q$; missing this fact led to incorrect
results in earlier publications. Expanding the distribution function
in angular harmonics in  momentum space, we mapped the Boltzmann
equation onto a system of recursion relations and found its
solution. 
In the particular case of a short-range correlated random
potential and/or random magnetic field
 the conductivity tensor takes a rather simple form, 
eq.(\ref{e10}). 
Application of our results 
 resolves an apparent inconsistency 
 with the interpretation of the experimental data on resonances of
composite fermions with surface acoustic waves. Using our formulas
 with experimental values of frequency and the CF effective
mass $m^*=0.8 m_e$, as extracted from the SdH measurements, we obtained the
curves (Fig.1) which describe well the experimentally observed
resonances (Fig.1 of Ref.\cite{willett2}). Our results are also to be
used in all the cases when the propagator of the gauge field (coupled
to CF's) at finite $\omega$ and $q$ is considered. This is important, in
particular, when one discusses the damping of SdH oscillations at
finite temperatures due to dynamic fluctuations of the gauge field
\cite{aamw,lee}. 

We are grateful to P.A.Lee for an interesting discussion which
motivated this work.
This project was supported by SFB 195 der
Deutschen Forschungsgemeinschaft and by the German-Israeli Foundation.


\begin{references}
\bibitem{note2} Note that taking the limit $\omega_c\to 0$
in Eqs.(\ref{e10}) for the model I is not  trivial, since
the arguments and the indices of the Bessel functions diverge in this
limit. It is more convenient to use the integral representation
(\ref{e52}) in this case, which in contrast simplifies in the
limit $\omega_c\to 0$.
\bibitem{note} Note that the condition $\omega_c\tau\gg 1$ is not in
conflict with the requirement $\omega_c\tau_s\ll 1$, which guarantees
that the magnetooscillations are weak and thus excludes the quantum
Hall regime, since $\tau\gg\tau_s$ in the presense of random magnetic
field scattering \cite{amw,aamw}.
\end{references}
\end{document}